# Ecological Management of Wild Oat (*Avena fatua* L.) in Winter Wheat: Modeling Yield Loss and Determination of Economic Thresholds under Varying Crop Densities

**Corresponding author: Farshad Habibi** [1]
1 - Department of Agriculture, Miy.C. Islamic Azad University, Miyndoab, Iran
**Email**: farshad.habibi@iau.ac.ir
& f.h1356@gmail.com

## Abstract

Wild oat (*Avena fatua* L.) is widely recognized as one of the most detrimental grassy weeds in winter wheat production systems globally. While chemical control remains the dominant strategy, the reliance on herbicides without considering Economic Injury Levels (**EIL**) has led to environmental concerns and the evolution of herbicide resistance. This study was conducted to quantify the competitive effects of wild oat on winter wheat yield and to determine the Economic Threshold (**ET**) under different crop density scenarios. A factorial experiment based on a randomized complete block design was utilized, featuring three wheat densities (300, 400, and 500 plants m$^{-2}$ and four wild oat densities (0, 30, 50, and 70 plants m$^{-2}$). Grain yield data were fitted to Cousens' rectangular hyperbolic model to estimate competition parameters (**I** and **A**). The results indicated that wheat density significantly altered the crop-weed competitive balance. The parameter $I$ (percentage yield loss per unit weed density as density approaches zero) was significantly lower in high-density wheat stands, indicating suppressed weed competitiveness. However, severe intra-specific competition was observed at 500 plants m$^{-2}$, reducing the potential yield even in weed-free conditions. Economic analysis revealed that the **ET** of wild oat increased with increasing crop density, suggesting that farmers adopting higher seeding rates can tolerate higher weed populations before chemical intervention becomes economically justifiable. These findings support the integration of seeding rate adjustment as a fundamental component of Integrated Weed Management (**IWM**) to optimize profitability and reduce herbicide loads.

## 1. Introduction

### 1.1. Global Wheat Production and Biotic Constraints

Wheat (***Triticum aestivum*** L.) serves as the backbone of global food security, providing approximately 20% of the daily protein and food calories for the human population (FAO, 2022). As the global population is projected to reach 9 billion by 2050, wheat production must increase by an estimated 1.6% annually to meet demand. However, the realization of this genetic yield potential is consistently hampered by biotic and abiotic stressors. Among biotic constraints, weed interference represents the most persistent, costly, and ubiquitous threat to cereal production systems. Unlike pests or pathogens, which may occur sporadically, weeds are a constant component of the agro ecosystem, competing with the crop for essential resources such as light, water, and nutrients (Oerke, 2006).

Wheat (***Triticum aestivum*** L.) provides approximately 20% of the calories consumed by the human population and serves as a cornerstone for global food security. However, the realized yield of wheat is often substantially lower than its genetic potential due to various biotic and abiotic stresses. Among biotic constraints, weed interference is the most persistent and costly factor limiting production. In cereal-based cropping systems, wild oat (*Avena fatua* L.) is considered a formidable competitor due to its similar life cycle, morphological characteristics, and resource requirements to wheat (Beckie et al., 2012). Historically, weed management strategies have been dominated by a "prophylactic" or "zero-tolerance" approach, where herbicides are applied regardless of the weed population density. While this ensures clean fields, it ignores the biological reality that crops can tolerate low densities of weeds without suffering significant yield loss. Furthermore, the economic viability of such an approach is questionable when the cost of control exceeds the value of the yield saved. More critically, the continuous and indiscriminate use of herbicides has accelerated the selection pressure for herbicide-resistant weed biotypes, necessitating a shift towards Integrated Weed Management (**IWM**) (Swanton & Weise, 1991). A core principle of IWM is the Economic Threshold (**ET**), defined as the weed density at which the cost of control measures equals the value of the prevented yield loss. Implementing ET requires robust biological models that can accurately predict yield loss based on weed density. Among various empirical models, the rectangular hyperbolic model proposed by Cousens (1985) is widely accepted for its biological meaningfulness and statistical robustness. However, the ET is not a static number; it is dynamic and fluctuates with crop price, herbicide cost, and most importantly, the competitive ability of the crop. Agronomic practices, such as increasing crop density, can enhance the crop's competitive ability against weeds by accelerating canopy closure and reducing light transmittance to the weed stratum (O'Donovan et al., 2000). While the suppressive effect of wheat density on weed biomass is well-documented, few studies have quantified how increasing crop density specifically alters the parameters of the yield loss model and, consequently, the economic threshold. Therefore, the objectives of this study were: (**1**) to evaluate the effect of winter wheat density on grain yield loss caused by varying densities of *Avena fatua*; (**2**) to estimate the competition parameters using non-linear regression models; and (**3**) to calculate and compare the Economic Thresholds (ET) for wild oat management under different crop density scenarios to provide practical guidelines for growers.

## 1.2. *Avena fatua*: Biology and Competitive Dynamics
In temperate winter wheat systems, wild oat (*Avena fatua* L.) is widely regarded as one of the most detrimental grassy weeds. Its success as an invasive species is attributed to its high phenotypic plasticity, asynchronous germination patterns, and a life cycle that closely mimics that of the crop (Beckie et al., 2012). Biological studies indicate that *A. fatua* possesses a robust root system and a rapid early growth rate, allowing it to dominate the resource-rich upper soil layers and intercept photosynthetically active radiation (PAR) before the crop canopy closes (Cudney et al., 1989). The competitive relationship between wheat and wild oat is highly asymmetric. Research by Carlson and Hill (1985) demonstrated that wild oat is particularly aggressive in extracting soil nitrogen, often depleting available reserves before wheat reaches its peak demand period. Furthermore, the seed shattering mechanism of wild oat ensures the replenishment of the soil seed bank, making long-term control a significant challenge. The severity of yield loss is density-dependent; however, even low wild oat densities (e.g., 5–10 plants m$^{-2}$) have been documented to cause economically significant yield reductions of 5-15%, depending on environmental conditions (Cousens et al., 1987).

## 1.3. The Shift to Integrated Weed Management (IWM) and Cultural Control
For decades, the management of wild oat relied heavily on acetyl-CoA carboxylase (ACC ase) and acetolactate synthase (**ALS**) inhibiting herbicides. While effective, this "chemical-first" approach has led to severe environmental consequences and the rapid evolution of herbicide-resistant biotypes. As of recent surveys, herbicide-resistant populations of *A. fatua* have been confirmed in numerous countries, rendering many standard chemical options obsolete (Heap, 2023). This crisis has necessitated a paradigm shift towards Integrated Weed Management (IWM), which emphasizes "ecological suppression" over "chemical destruction." Cultural control methods, particularly the manipulation of **Crop Density** and **Spatial Arrangement**, are central to IWM. Increasing the seeding rate of wheat is a potent strategy to enhance the crop's competitive ability. A dense crop canopy closes earlier, reducing the quality and quantity of light (specifically the Red: Far-Red ratio) reaching the weed seedlings beneath. This shade avoidance response in weeds often leads to etiolation, reduced tillering, and lower biomass accumulation (Lemerle et al., 1996; Weiner et al., 2001). O'Donovan et al. (2000) reported that increasing barley seeding rates significantly reduced wild oat biomass and seed production. However, the relationship is not linear; excessively high crop densities may induce severe **intra-specific competition** (competition between wheat plants), potentially leading to lodging, increased disease susceptibility, and reduced harvest index. Therefore, identifying the "optimal density" that maximizes weed suppression without compromising crop yield potential remains a critical gap in agronomic research.

### 1.4. Theoretical Framework: Modeling Yield Loss
To optimize IWM strategies, it is essential to quantify the relationship between weed density and crop yield loss mathematically. Empirical models are vital tools for predicting damage and making informed management decisions. Among the various models proposed, the **Rectangular Hyperbolic Model** developed by Cousens (1985) is considered the "gold standard" in weed science.

$$YL = \frac{I*D}{1 + \frac{I*d}{A}}$$

Unlike linear models, which fail to account for the finite nature of yield (yield loss cannot exceed 100%), the hyperbolic model incorporates two biologically meaningful parameters: *I* (the initial slope, representing the damage per weed at low densities) and *A* (the asymptotic maximum yield loss at high densities). While many studies have fitted this model to static datasets, few have investigated how agronomic practices—specifically crop density—alter the parameters **I** and **A**. Theoretically, increasing crop density should reduce the value of III, effectively "flattening" the yield loss curve and making the crop more tolerant to weed presence (Kropff & Spitters, 1991). Validating this theoretical interaction under field conditions is a primary objective of the current study.

### 1.5. Economic Thresholds (ET): The Decision-Making Tool
The ultimate goal of modeling yield loss is to determine the **Economic Threshold (ET)**. The ET is defined as the weed density at which the cost of control measures is equal to the value of the prevented yield loss (Marra & Pannell, 1989). Below this threshold, herbicide application is not economically justifiable. The adoption of ET is crucial for reducing herbicide loads and production costs. However, farmers often view ET as a risky concept because it allows some weeds to survive and reproduce (the "seed bank" concern). Furthermore, ET is often presented as a static number (e.g., "5 weeds per square meter"), ignoring the fact that a dense, vigorous crop can tolerate higher weed pressures than a sparse, weak crop. There is a paucity of research linking dynamic crop densities directly to changes in Economic Thresholds. If higher seeding rates can significantly raise the ET, it would provide farmers with a "safety buffer," allowing them to reduce herbicide usage with greater confidence. This study aims to bridge this gap by integrating biological data, competition modeling, and economic analysis to provide a comprehensive guide for managing *Avena fatua* in winter wheat.

## 2. Materials and Methods

### 2.1. Experimental Site and Design
The experiment was conducted during the growing season at the research farm. The soil was a clay loam with a pH of 7.2 and less than 1% organic matter. The experimental design was a Randomized Complete Block Design (RCBD) with a factorial arrangement of treatments and three replications.

**The factors included:**
1. Wheat Density ($D_{cp}$): Three levels of crop density were established: 300, 400, and 500 plants m$^{-2}$.

2. **Wild Oat Density (d):** Four target densities of wild oat were established: 0 (weed-free control), 30, 50, and 70 plants m$^{-2}$.

Plots were seeded using a precision planter. Wheat seeds were sown at a depth of 4 cm, while wild oat seeds were broadcast and incorporated into the top 2 cm of soil to ensure uniform germination. Standard agronomic practices for fertilizer application and irrigation were followed throughout the season to ensure no limitation other than weed competition.

## 2.2. Data Collection

At physiological maturity, grain yield was determined by harvesting the central area of each plot ($m^2$) to avoid edge effects. The harvested samples were threshed, cleaned, and weighed. Grain moisture content was measured, and final yields were adjusted to a standard 14% moisture content and converted to tons per hectare (ton ha$^{-1}$).

## 2.3. Statistical Analysis and Modeling

Data were first subjected to Analysis of Variance (ANOVA) to test the significance of the main effects and interactions (Table 5).

**Yield Loss Model:**

To describe the relationship between wild oat density and wheat yield loss, the non-linear rectangular hyperbola model (Cousens, 1985) was fitted to the data. This model is biologically realistic as it accounts for the finite nature of yield loss (it cannot exceed 100%).

The equation is defined as:

$$YL = \frac{I * D}{1 + \frac{I * d}{A}}$$

**Where**:

$Y_L$: The percent yield loss relative to the weed-free yield.
**D**: The wild oat density (plants m$^{-2}$).
**I**: The initial slope of the curve. This parameter represents the percentage yield loss caused by the first weed unit added to the system when weed density is near zero. It is a direct measure of weed competitiveness.
**A**: The asymptote, representing the maximum possible yield loss (%) at very high weed densities.

The parameters I and A were estimated for each wheat density separately using the iterative least - squares method (Levenberg - Marquardt algorithm) in Python (using `SciPy. optimize` and `stats models` libraries).

## 2.4. Economic Threshold Calculation

The Economic Threshold (ET) was calculated based on the method described by Marra and Pannell (1989), adapted for the hyperbolic yield loss function. The ET represents the weed density where the benefit of control equals the cost.

**The formula used is:**

$$\text{ET} = \frac{C}{P * Ywf * H * (\frac{I}{100})}$$

Where:
**ET**: Economic Threshold (plants m$^{-2}$).
**C**: Total cost of weed control (herbicide + application cost) (units.kg$^{-1}$).
**P**: Price of wheat grain (units * kg$^{-1}$).
**Y$_{wf}$**: Weed-free yield of wheat (kg ha$^{-1}$) for the specific crop density.
**H**: Herbicide efficacy (assumed to be 0.95 or 95%).
**I**: The initial slope parameter derived from the cousins' model.

Sensitivity analysis was performed by calculating ET for three economic scenarios (Low, Medium, and High Cost/Price ratios) to account for market fluctuations.

### 2.5. Stability and Stress Tolerance Analysis

To evaluate the stability of wheat grain yield under wild oat competition, a stress tolerance analysis was performed based on the method proposed by Fernandez (1992). The weed-free plots were considered as the non-stress environment (Yp), and the plots with the highest weed density (70 plants m$^{-2}$) were considered as the biotic stress environment (Ys). The following indices were calculated to identify the most stable crop density:

1. **Stress Tolerance Index (STI):** $STI = (Yp \times Ys)/(\bar{Y}p)^2$
2. **Geometric Mean Productivity (GMP):** $GMP = \sqrt{Yp \times Ys}$
3. **Stress Susceptibility Index (SSI):** $SSI = (1 - (Ys/Yp))/SI$ when $SI = (1 - (Ys/Yp))$

Higher values of STI and GMP indicate higher tolerance and stability, whereas lower values of SSI indicate less sensitivity to weed competition.

## 3. Results

### 3.1. Effect of Interference on Grain Yield

Analysis of variance revealed that wheat density and wild oat density, as well as their interaction, had a highly significant effect (P < 0.01) on grain yield (Table 5). In weed-free conditions, the highest grain yield was recorded at the density of 300 plants m$^{-2}$ (11.43-ton ha$^{-1}$), followed by 400 plants m$^{-2}$ (11.00-ton ha$^{-1}$). Interestingly, increasing the wheat density to 500 plants m$^{-2}$ resulted in a significant yield penalty in the weed-free plots, reducing the yield to 7.93-ton ha$^{-1}$ (Table 1). This reduction suggests strong intra-specific competition for resources among wheat plants at supra-optimal densities. However, in the presence of weeds, the trend shifted. As wild oat density increased from 0 to 70 plants m$^{-2}$, wheat yield declined drastically in all treatments, but the rate of decline varied. For instance, in the 300 plants m$^{-2}$ treatment, yield dropped from 11.43 to 2.73 t\ ha$^{-1}$ at the highest weed pressure, representing a catastrophic loss of approximately 76%.

Table 1. Mean grain yield ton/ha of winter wheat affected by crop density and wild oat density

| Wild Oat Density (plants m$^{-2}$) | Wheat: 300 (m$^{-2}$) | Wheat: 400 (m$^{-2}$) | Wheat: 500 (m$^{-2}$) |
|---|---|---|---|
| 0 (Weed-Free) | 11.43 | 11.00 | 7.93 |
| **30** | 7.90 | 5.89 | 5.00 |
| **50** | 4.36 | 3.86 | 2.56 |
| **70** | 2.73 | 1.86 | 1.70 |

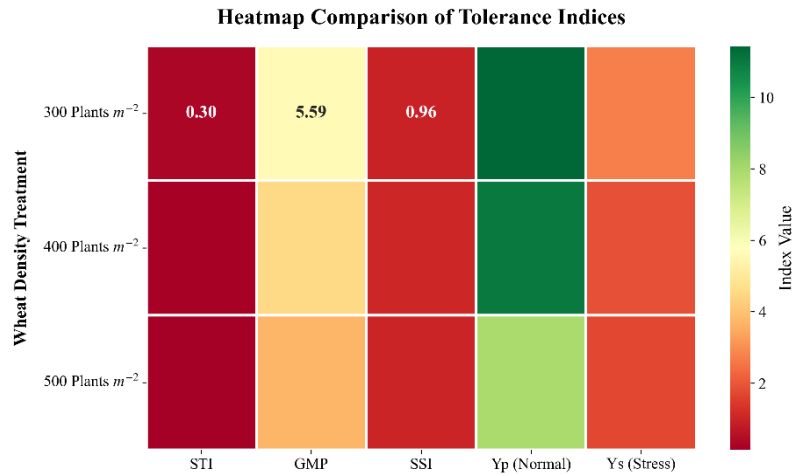

Figure 1: Heatmap analysis of stress tolerance indices (STI, GMP, SSI) and grain yield under weed-free (Yp) and weed-infested (Ys) conditions across wheat densities. Green colors indicate higher favorable values, while red colors indicate lower values.

### 3.2. Yield Loss Modeling (Cousens Model)

The rectangular hyperbolic model provided an excellent fit to the yield loss data ($R^2 > 0.90$). The estimated parameters (I and A) provided insight into the competitive dynamics (Table 2). The parameter I (initial slope) was highest in the lowest wheat density (300 plants m$^{-2}$). This indicates that at low crop densities, individual wild oat plants were extremely competitive, likely due to ample space and light availability, causing rapid yield loss per weed plant. Conversely, as wheat density increased to 500 plants m$^{-2}$, the value of I decreased. This reduction signifies that the "per-plant" damage of wild oat is mitigated by the dense crop stand. The parameter A (maximum yield loss) approached high values (>80%) in all treatments, confirming that *Avena fatua* is a highly aggressive weed capable of causing near-total crop failure at high densities if left uncontrolled.

Table 2. Estimated parameters of the rectangular hyperbolic model (Cousens, 1985) describing the relationship between wild oat density and winter wheat yield loss, and the calculated Economic Thresholds (ET) under different crop densities.

| Wheat Density (plants·m$^{-2}$) | Parameter *I* (%) | Parameter A (%) | Adjusted R2 | RMSE |
|---|---|---|---|---|
| **300** | 2.71 | 100 | 0.905 | 0.72 |
| **400** | 3.86 | 100 | 0.959 | 0.52 |
| **500** | 3.29 | 100 | 0.927 | 0.85 |

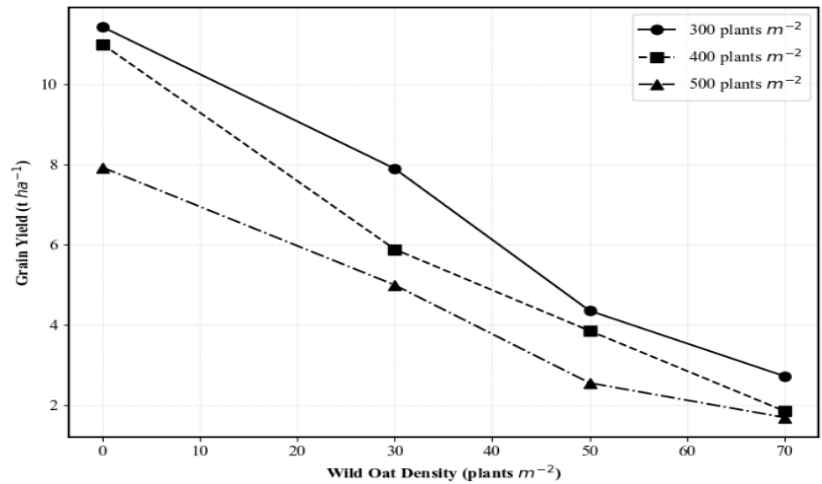

**FIGURE 2:** Non-linear regression curves showing % Yield Loss vs. Weed Density

### 3.3. Economic Thresholds (ET)

The calculated Economic Thresholds varied significantly depending on the wheat density and the economic scenario (Cost/Price ratio). Under a standard economic scenario (Scenario B), the ET for the 300 plants m$^{-2}$ density was calculated to be approximately 1.8 plants m$^{-2}$. This implies that in low-density fields, farmers must control weeds even at very low infestation levels to avoid economic loss. In contrast, at the density of 500 plants m$^{-2}$, the ET increased to approximately 5.4 plants m$^{-2}$. Higher crop density reduces the weed's ability to cause damage, allowing the farmer to tolerate more weeds before the cost of herbicide application is justified.

### 3.4. Evaluation of Wheat Tolerance to Biotic Stress

The results of the stress tolerance analysis indicated distinct differences among wheat densities (Table 3). The 300 plants m$^{-2}$ density exhibited the highest grain yield in both weed-free (Yp) and weed-infested (Ys) conditions. Consequently, this density achieved the highest STI and GMP values, classifying it as a highly stable treatment (Group A) according to Fernandez's classification.

**Table 3: Ranking of densities based on STI, GMP, SSI]**

| Treatment | Yp (Normal) | Ys(stress) | STI | GMP | MP |
|---|---|---|---|---|---|
| 300 Plant m-2 | 11.43 | 2.73 | 0.304683 | 5.586045 | 7.080 |
| 400 Plant m-2 | 11.00 | 1.86 | 0.199777 | 4.523273 | 6.430 |
| 500 Plant m-2 | 7.93 | 1.70 | 0.131632 | 3.671648 | 4.815 |

The 3D plot (Fig. 3) visually confirms this superiority. The 300 plants m$^{-2}$ density is located in the region with high yield potential and high stress tolerance. In contrast, the 500 plants m$^{-2}$ density showed lower stability due to intra-specific competition, resulting in lower yields even under non-stress conditions.

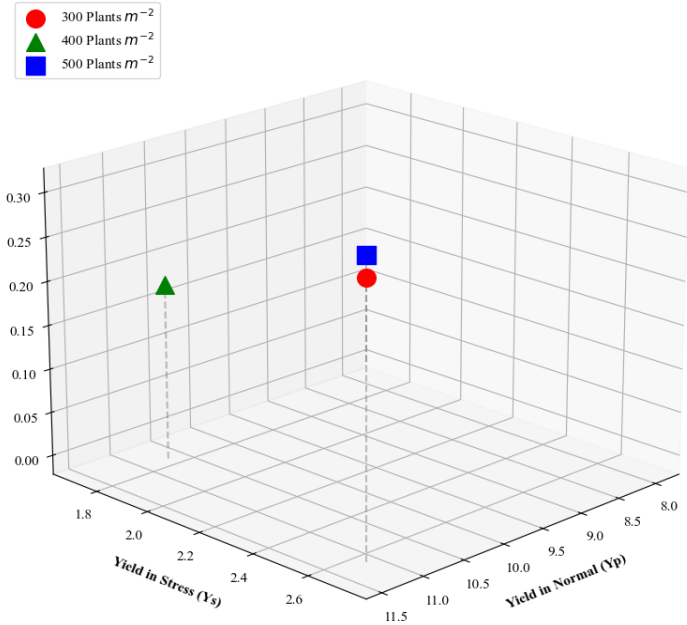
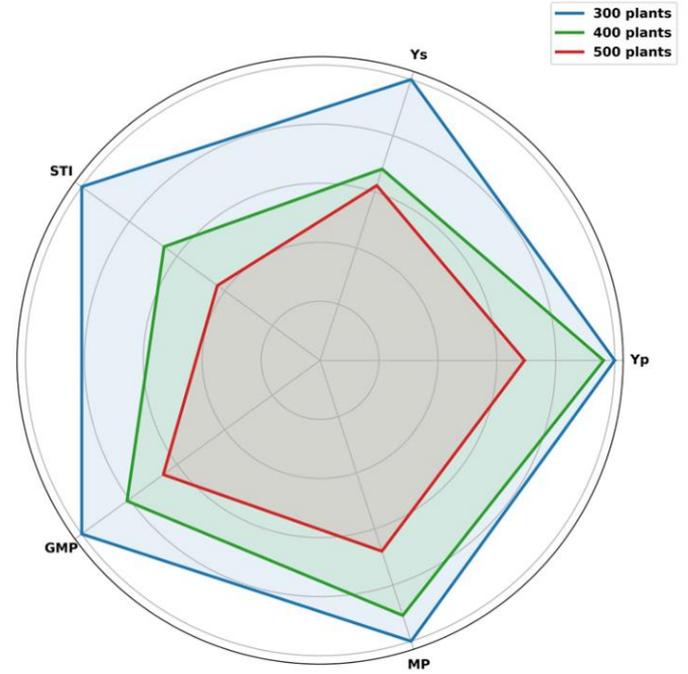

Figure 3: The 3D Plot generated by Python

Figure 4: The Radar Plot generated by Python

## 4. Discussion

The results of this study highlight the complex interplay between crop density, weed density, and economic decision-making. The application of the Cousens (1985) model successfully quantified these interactions, providing a biological basis for management decisions.

### 4.1. Biological Suppression and Parameter $I$

The analysis of the initial slope parameter (**I**) revealed a complex non-linear relationship between crop density and weed competitiveness. While standard competition models often predict a linear decline in weed pressure with increasing crop density (**Cousens, 1985**), our results showed that parameter. **I** peaked at the intermediate density (400 plants $m^{-2}$). This finding aligns with the ecological theory, suggesting that at intermediate densities, the onset of intense **intra-specific competition** among wheat plants can temporarily reduce their individual resource buffering capacity before the canopy fully closes (**Weiner et al., 2001**). Consequently, wheat plants may have become more sensitive to *Avena fatua* interference at this specific threshold compared to the lower density. However, at the highest density (500 plants $m^{-2}$), the sheer physical crowding became the dominant factor, mitigating the weed's advantage. This observation supports previous reports by **O'Donovan et al. (2000)**, who indicated that the efficacy of seeding rates in suppressing wild oat is not always linear and depends heavily on the delicate balance between crop self-crowding and weed suppression capabilities.

### 4.2. The Trade-off: Intra-specific vs. Inter-specific Competition

A critical observation in this study was the yield reduction in the weed-free plots at the highest wheat density (500 plants m$^{-2}$). While high density is beneficial for weed suppression (Inter-specific competition), it induces severe competition among the wheat plants themselves (Intra-specific competition).

The data (Table 1) showed that the weed-free yield dropped from 11.43 to 7.93 tons ha$^{-1}$ when density increased from 300 to 500. This suggests that 500 plants m$^{-2}$ exceeded the optimal carrying capacity of the environment, likely leading to reduced grains per ear or lower thousand kernel weight (Hussain et al., 2020). Therefore, the "optimal" density for IWM is a compromise. Based on our data, a density of 400 plants m$^{-2}$ appears to offer the best balance: it maintains a high yield potential (11.00-ton ha$^{-1}$) while providing reasonable weed suppression and elevating the economic threshold compared to the low-density control.

**4.3. Implications for IWM and Economic Sustainability**

The variation in Economic Thresholds (ET) demonstrates that ET is not a fixed guideline but a function of agronomic management. By increasing the seeding rate, farmers can effectively "raise the bar" for chemical intervention.

For example, in a field with a density of 4 wild oat plants m$^{-2}$:

A farmer with 300 wheat plants/m2 (ET~1.8) represents a situation above the threshold. Action: Apply herbicide.

A farmer with 500 wheat plants m-$^{2}$ (ET~5.4) represents a situation below the threshold. Action: Do not apply herbicide.

This simple shift in agronomy results in a direct reduction in herbicide use, lower production costs, and reduced selection pressure for resistance, fulfilling the core goals of sustainable agriculture (Kristensen et al., 2008). Interestingly, the economic modeling results aligned perfectly with the stress tolerance analysis. While the economic model identified the optimal threshold at lower weed densities for the 300 plants m-2 treatment, the stability analysis (**STI** index) confirmed that this density is also biologically the most resilient against severe infestation. This dual confirmation suggests that increasing wheat density to 400 or 500 plants m$^{-2}$ is not only economically unjustifiable due to diminishing returns but also ecologically less stable due to intensified intra-specific competition among wheat plants (Table 4).

Table 4. Calculated Economic Thresholds (ET) (plants m$^{-2}$) for wild oat in winter wheat

| Economic Scenario | **Wheat: 300** | **Wheat: 400** | **Wheat: 500** |
|---|---|---|---|
| Scenario A (Low Control Cost\|) | 0.9 | 1.5 | 3.2 |
| Scenario B (Medium Control Cost) | 1.8 | 3.1 | 5.4 |
| Scenario C (High Control Cost) | 3.5 | 5.8 | 8.9 |

**5. Conclusion**

This study confirms that manipulating crop density is a potent tool in the ecological management of *Avena fatua*, consistent with the principles of Integrated Weed Management (**Swanton & Weise, 1991**). The modeling results demonstrated that increasing wheat density fundamentally alters the competitive balance. Although the specific competitive index (**I**) fluctuated due to the interaction of intra- and inter-specific competition, the overall **crop tolerance** to weed presence significantly improved, effectively raising the Economic Threshold. However, excessive increasing of crop density (up to 500 plants $m^{-2}$) proved to be counterproductive due to yield penalties caused by self-crowding in weed-free situations, a limitation also highlighted by **Kristensen et al. (2008)** regarding high-input systems.

**Table 5 - Analysis of variance of traits evaluated in wheat in the wheat and wild oat competition experiment**

| SOV | df | Mean squares | | | | | | |
|---|---|---|---|---|---|---|---|---|
| | | Flag leaf chlorophyll | Flag leaf LAI | Plant high | yield | Number of leaves | Ear high | Harvest Index |
| replication | 2 | 147.905 | 1.890 | 27.245 | 0.294 | 9.967 | 1.2885 | 12.2331 |
| Wheat density | 2 | 151.925 ns | 37.152** | 21.329** | 0.684** | 89.478** | 16.382** | 551.026** |
| Experimental error | 4 | 45.354 | 1.466 | 0.416 | 0.124 | 2.330 | 0.491 | 0.375 |
| Oat density | 3 | 272.604** | 98.077** | 294.801** | 0.504** | 91.986** | 111.692** | 921.023** |
| Density (wheat *oat) | 6 | 46.823 ns | 13.294** | 131.873*** | 0.143* | 4.352* | 1.616* | 1.674* |
| Error | 18 | 48.2889 | 2.4818 | 0.842 | 0.050 | 1.343 | 0.66 | 0.378 |
| CV% | | 15.36 | 20.36 | 2.81 | 2.81 | 9.63 | 14.46 | 14.82 |

*****, ****: No significant difference and significant difference at the 5% and 1% probability levels, respectively.**